\documentclass[11pt]{article}%[12pt]{iopart}
\input{epsf}
\usepackage{amsmath}
\usepackage{amssymb}

\usepackage{booktabs}

\DeclareMathOperator{\PP}{\mathbb{P}}%BlackBoardP
\DeclareMathOperator{\BPP}{\bar{\mathbb{P}}}%BaredBlackBoardP

\begin{document} 
\title{Notes on Cubic and Quartic Light-Front Kinematics}       
\author{Anders K. H. Bengtsson\footnote{e-mail: anders.bengtsson@hb.se. Work supported by the Research and Education Board at the University of Bor{\aa}s.}}

\maketitle
\begin{center}School of Textiles, Engineering and Economics\break University of Bor{\aa}s\break All\' egatan 1, SE-50190 Bor\aa s, Sweden.\end{center}

\begin{abstract}
The purpose of this short note is to collect a set of formulas pertaining to momentum kinematics for higher spin light-front vertices. At least one of the formulas seems to be previuosly unknown.

\end{abstract}

%\pagebreak
%\tableofcontents 
%\pagebreak

%=============================================
\section{Introduction}\label{sec:Introduction}
%=============================================
In a previous paper \cite{AKHB2012a} I have tried to systematise the structure of four dimensional higher spin light-front vertices. What was lacking there was certain \emph{recombination formulas}. Formulas that are likely to be crucial -- at least technically -- in an attempt carry through a computation the full quartic vertex (if it exists). This deficiency is filled by the present notes.

%=============================================
\section{Trivia}\label{sec:Introduction}
%=============================================
I'm are working on a $D=4$ Minkowski light front with {\tiny $-+++$} metric where the transverse directions are written in complex notation. Explicitly we have the following basic formulas.
\begin{align}\label{eq:BasicDefs1}
\gamma\equiv p^+&=\frac{1}{\sqrt{2}}\big(p^0+p^3\big)\quad\quad\quad p^-=\frac{1}{\sqrt{2}}\big(p^0-p^3\big)\\\label{eq:BasicDefs2}
p&=\frac{1}{\sqrt{2}}\big(p^1+ip^2\big)\quad\quad\quad \bar{p}=\frac{1}{\sqrt{2}}\big(p^1-ip^2\big)
\end{align}
In this notation, the metric becomes $A_\mu B^\mu=A\bar{B}+B\bar{A}-A^+B^--A^-B^+$ and we have the mass-shell condition
\begin{equation}\label{eq:MassShell}
p^+p^--p\bar{p}=0\quad\Rightarrow\quad h\equiv p^-=\frac{p\bar{p}}{p^-}=\frac{p\bar{p}}{\gamma}
\end{equation}
In any $n$-point vertex we have momentum conservation (taking all momenta as outgoing)
\begin{align}
\sum_{r=1}^n p_r=\sum_{r=1}^n \bar{p}_r=\sum_{r=1}^n \gamma_r=0
\end{align}
But note that the free hamiltonians
\begin{equation}
h_r=p_r^-=\frac{p_r\bar{p}_r}{\gamma_r}
\end{equation}
do not sum to zero. This is our starting point.

%======================================================
\section{Cubic recombination}\label{sec:Cubicrecombination}
%======================================================
For the cubic vertices we first introduce the transverse momentum variables
\begin{equation}\label{eq:DefBlackboardMomentumN3}
\PP=-\frac{1}{3}\sum_{r=1}^3\widetilde\gamma_r p_r\quad\text{with}\quad \widetilde\gamma_r=\gamma_{r+1}-\gamma_{r+2}
\end{equation}
and correspondingly for the $\BPP$. Explicitly we have
\begin{equation}\label{eq:DefBlackboardMomentumExplicitN3}
\PP=\frac{1}{3}\big(\PP_{12}+\PP_{23}+\PP_{31}\big)
\end{equation}
with 
\begin{equation}\label{eq:DefBlackboardMomentumExplicitDEFN3}
\PP_{12}=\gamma_1 p_2-\gamma_2 p_1,\quad \PP_{23}=\gamma_2 p_3-\gamma_3 p_2,\quad\PP_{31}=\gamma_3 p_1-\gamma_1 p_3
\end{equation}
Momentum conservation yields $\PP_{12}=\PP_{23}=\PP_{31}$. Corresponding formulas hold for the $\BPP$.

%-------------------------------------------------------------------
\subsection{Sum over the hamiltonians}\label{subsec:SumHamiltonians}
Rather than summing to zero, we get for the sum of the three external free hamiltonians
\begin{equation}\label{eq:CubicSumHamiltonians}
\sum_{r=1}^3 h_r=-\frac{\PP\BPP}{\gamma_1\gamma_2\gamma_3}
\end{equation}
This formula is a special case of a general cubic recombination formula.

%--------------------------------------------------------------------
\subsection{Cubic recombination formula}\label{subsec:Cibicrecombination}
Let $c_r$ for $r=1,2,3$ be arbitrary coefficients and consider the linear combination
\begin{equation}\label{eq:CubicLinearCombination}
\sum_{r=1}^3 c_rp_r
\end{equation}
Since the linear space spanned by the transverse momenta $\{p_1,p_2,p_3\}$ is really two-dimensional, due to momentum conservation, one would expect to be able to rewrite \eqref{eq:CubicLinearCombination} as linear combination over two independent basis vectors. This is indeed possible, and the formula reads as follows \cite{LindenCubicResum}.
\begin{equation}\label{eq:recombinationFormulaCubic}
\sum_{r=1}^3 c_rp_r=\frac{1}{3}\Big(\sum_{r=1}^3 c_r\gamma_r\Big)\Big(\sum_{s=1}^3\frac{p_s}{\gamma_s}\Big)+\frac{1}{3\gamma_1\gamma_2\gamma_3}\Big(\sum_{r=1}^3 c_r\gamma_r\widetilde{\gamma}_r\Big)\PP
\end{equation}
where we interpret the new basis vectors as 
\begin{equation}\label{eq:NewCubicBasis}
\sum_{s=1}^3\frac{p_s}{\gamma_s}\quad\text{and}\quad \PP
\end{equation}
It is easy to see that the formula \eqref{eq:CubicSumHamiltonians} is a special case of this with $c_r=\bar{p}_r/\gamma_r$. The first term on the right is zero by momentum conservation while the second becomes
\begin{equation*}
\frac{1}{3\gamma_1\gamma_2\gamma_3}\Big(\sum_{r=1}^3 \frac{\bar{p}_r}{\gamma_r}\gamma_r\widetilde{\gamma}_r\Big)\PP=\frac{1}{3\gamma_1\gamma_2\gamma_3}\Big(\sum_{r=1}^3 \bar{p}_r\widetilde{\gamma}_r\Big)\PP=-\frac{\BPP}{\gamma_1\gamma_2\gamma_3}\PP
\end{equation*}

The recombination formula is most useful precisely in cases when the first term on the right is zero. This is what happens when the coefficients are such that the sum $\sum_{r=1}^n c_r\gamma_r$ is zero. 

Focusing on the coefficient of the second term on the right we note that it can be written as
\begin{equation}\label{eq:recombinationFormulaCubic2ndterm}
\frac{1}{\gamma_1\gamma_2\gamma_3}\Big(\sum_{r=1}^3 c_r\gamma_r\widetilde{\gamma}_r\Big)=c_1\left(\frac{1}{\gamma_3}-\frac{1}{\gamma_2}\right)+c_2\left(\frac{1}{\gamma_1}-\frac{1}{\gamma_3}\right)+c_3\left(\frac{1}{\gamma_2}-\frac{1}{\gamma_1}\right)
\end{equation}
This is the form that will generalise to higher order and the quartic in particular.

%==========================================================
\section{Quartic recombination}\label{sec:Quarticrecombination}
%==========================================================
A clue to the quartic generalisation of the recombination formula comes from the spinor helicity formalism (reviewed in for instance \cite{Dixon2013Review,FengLuo2012Review,ElvangHuangBook}) which has, as has been noted \cite{Ananth2012un}, certain similarities with light-front notation. However, we will exploit precisely a point where the analogy does not hold up.

%---------------------------------------------------------------------------------------
\subsection{Spinor helicity -- light front correspondence}\label{subsec:SpinorHelicityLF}
Contracting the momenta with Pauli matrices we get
\begin{equation}\label{eq:Pslash}
p_{a\dot a}=p_\mu\sigma^\mu_{a\dot a}=\sqrt{2}\begin{pmatrix}-p^-&\bar p\cr p &-p^+\end{pmatrix}=/\text{on shell}/=\sqrt{2}\begin{pmatrix}-\frac{p\bar p}{\gamma}&\bar p\cr p &-\gamma\end{pmatrix}
\end{equation}
The determinant of this two-by-two matrix is zero. It can be written as a product of two-component angle and bracket spinors
\begin{equation}\label{eq:LFSpinors}
\begin{split}
|p]_a=\frac{\sqrt[4]2}{\sqrt\gamma}\begin{pmatrix}\bar p\\-\gamma\end{pmatrix}\quad\text{and}\quad\langle p|_{\dot a}=\frac{\sqrt[4]2}{\sqrt\gamma}\begin{pmatrix}p\,,&-\gamma\end{pmatrix}
\end{split}
\end{equation}
When two such spinors for different momenta $p_1$ and $p_2$ are bracketed, we get
\begin{equation}\label{eq:AngleBracketP}
\begin{split}
\langle 1 2\rangle&\equiv\langle p_1 p_2\rangle=\langle p_1|_{\dot a}| p_2\rangle^{\dot a}=\frac{\sqrt[4]{2}}{\sqrt{\gamma_1}}(p_1,
-\gamma_1)\frac{\sqrt[4]{2}}{\sqrt{\gamma_2}}\begin{pmatrix}-\gamma_2\cr
-p_2\end{pmatrix}=\frac{\sqrt{2}}{\sqrt{\gamma_1\gamma_2}}\mathbb{P}_{12}\\
[1 2]&\equiv[p_1 p_2]=[p_1|^a|p_2]_a=-\frac{\sqrt{2}}{\sqrt{\gamma_1\gamma_2}}\mathbb{\bar{P}}_{12}
\end{split}
\end{equation}
In this formalism, momentum conservation takes the form
\begin{equation}\label{eq:IdentitiesMomCons}
\sum_{i=0}^n|i\rangle[i|=0\quad\Rightarrow\quad\sum_{i=0}^n\langle p\,i\rangle[i\,q]=0
\end{equation}
When expressed in terms of the light-front form, we get
\begin{equation}\label{eq:momnoncons}
\sum_{i=0}^n|i\rangle[i|=\sum_{i=0}^n\frac{\sqrt[4]2}{\sqrt\gamma_i}\begin{pmatrix}\gamma_i\\p_i\end{pmatrix}\frac{\sqrt[4]2}{\sqrt\gamma_i}\begin{pmatrix}\gamma_i\,,&\bar p_i\end{pmatrix}=\sqrt{2}\sum_{i=0}^n\begin{pmatrix}\gamma_i&\bar p_i\\p_i&\frac{p_i\bar p_i}{\gamma_i}\end{pmatrix}
\end{equation}
The terms $\frac{p_i\bar p_i}{\gamma_i}$ do not sum to zero. Instead, for $n=3$, formula \eqref{eq:CubicSumHamiltonians} is reproduced
\begin{equation}\label{eq:n3noncons}
\begin{split}
\sum_{i=0}^3\frac{p_i\bar p_i}{\gamma_i}=-\frac{\PP\BPP}{\gamma_1\gamma_2\gamma_3}
\end{split}
\end{equation}
However, for the moment forgetting that all $\PP_{ij}$ are essentially equal for $n=3$ we actually get six ways to express the sum
\begin{equation}\label{eq:n3noncons}
\begin{split}
\sum_{i=0}^3\frac{p_i\bar p_i}{\gamma_i}&=-\frac{\PP_{12}\BPP_{23}}{\gamma_1\gamma_2\gamma_3}=-\frac{\PP_{23}\BPP_{31}}{\gamma_2\gamma_3\gamma_1}=-\frac{\PP_{31}\BPP_{12}}{\gamma_3\gamma_1\gamma_2}\\
&=-\frac{\PP_{13}\BPP_{32}}{\gamma_1\gamma_3\gamma_2}=-\frac{\PP_{21}\BPP_{13}}{\gamma_2\gamma_1\gamma_3}=-\frac{\PP_{32}\BPP_{21}}{\gamma_3\gamma_2\gamma_1}
\end{split}
\end{equation}
The corresponding result for $n=4$ will be used to derive the quartic recombination formula.

%-----------------------------------------------------------------------------
\subsection{Quartic transverse momenta}\label{subsec:QuarticTransverseMomenta}
For $n=4$ there are six objects each of $\PP_{ij}$ and $\BPP_{ij}$. Explicitly we have the set
\begin{equation*}
\PP_{12},\PP_{13},\PP_{14},\PP_{23},\PP_{24},\PP_{34}
\end{equation*}
defined through $\PP_{ij}=\gamma_ip_j-\gamma_jp_i$ and likewise for the $\BPP_{ij}$.

Of these, only two $\PP_{ij}$ and two $\BPP_{ij}$ are independent due to momentum conservation \cite{Metsaev2005ar}. Actually, there is one purely algebraic equation linking the variables
\begin{equation}\label{eq:AlgebraicEquation}
\frac{\PP_{12}}{\gamma_1\gamma_2}+\frac{\PP_{23}}{\gamma_2\gamma_3}+\frac{\PP_{34}}{\gamma_3\gamma_4}+\frac{\PP_{41}}{\gamma_4\gamma_1}=0
\end{equation}
that follows from the definition. Through momentum conservation, one finds the two sets of equations
\begin{equation}\label{eq:MomConsEqs3}
\begin{cases}
\PP_{12}+\PP_{13}+\PP_{14}=0\\
\PP_{12}-\PP_{23}-\PP_{24}=0\\
\PP_{13}+\PP_{23}-\PP_{34}=0\\
\PP_{14}+\PP_{24}+\PP_{34}=0
\end{cases}
\end{equation}
and
\begin{equation}\label{eq:MomConsEqs4}
\begin{cases}
\PP_{12}+\PP_{13}-\PP_{24}-\PP_{34}=0\\
\PP_{12}+\PP_{14}+\PP_{23}+\PP_{34}=0\\
\PP_{13}+\PP_{14}+\PP_{23}+\PP_{24}=0
\end{cases}
\end{equation}
Of these last seven equations only three are linearly independent. All in all, equations \eqref{eq:AlgebraicEquation}, \eqref{eq:MomConsEqs3} and \eqref{eq:MomConsEqs4} can be solved in terms of different choices of two $\PP_{ij}$. 

It makes some sense to call $\PP_{12}$ and $\PP_{34}$ $\mathbf{s}$-channel variables, $\PP_{14}$ and $\PP_{23}$ $\mathbf{t}$-channel variables and $\PP_{13}$ and $\PP_{24}$ $\mathbf{u}$-channel variables. With this terminology, we get

%§§§§§§§§§
\paragraph{$\mathbf{s}$-channel variables}
\begin{equation}\label{eq:SchannelSolution}
\begin{cases}
\PP_{24}=\frac{1}{\gamma_3+\gamma_4}\left(\gamma_4\PP_{12}+\gamma_2\PP_{34}\right)\\
\PP_{13}=\frac{1}{\gamma_3+\gamma_4}\left(-\gamma_3\PP_{12}-\gamma_1\PP_{34}\right)\\
\PP_{14}=\frac{1}{\gamma_3+\gamma_4}\left(-\gamma_4\PP_{12}+\gamma_1\PP_{34}\right)\\
\PP_{23}=\frac{1}{\gamma_3+\gamma_4}\left(\gamma_3\PP_{12}-\gamma_2\PP_{34}\right)
\end{cases}
\end{equation}
%

%§§§§§§§§§
\paragraph{$\mathbf{t}$-channel variables}
\begin{equation}\label{eq:TchannelSolution}
\begin{cases}
\PP_{24}=\frac{1}{\gamma_2+\gamma_3}\left(-\gamma_2\PP_{14}+\gamma_4\PP_{23}\right)\\
\PP_{13}=\frac{1}{\gamma_2+\gamma_3}\left(-\gamma_3\PP_{14}+\gamma_1\PP_{23}\right)\\
\PP_{12}=\frac{1}{\gamma_2+\gamma_3}\left(-\gamma_2\PP_{14}-\gamma_1\PP_{23}\right)\\
\PP_{34}=\frac{1}{\gamma_2+\gamma_3}\left(-\gamma_3\PP_{14}-\gamma_4\PP_{23}\right)
\end{cases}
\end{equation}
%

%§§§§§§§§§
\paragraph{$\mathbf{u}$-channel variables}
\begin{equation}\label{eq:UchannelSolution}
\begin{cases}
\PP_{12}=\frac{1}{\gamma_1+\gamma_3}\left(\gamma_2\PP_{13}+\gamma_1\PP_{24}\right)\\
\PP_{34}=\frac{1}{\gamma_1+\gamma_3}\left(-\gamma_4\PP_{13}-\gamma_3\PP_{24}\right)\\
\PP_{14}=\frac{1}{\gamma_1+\gamma_3}\left(\gamma_4\PP_{13}-\gamma_1\PP_{24}\right)\\
\PP_{23}=\frac{1}{\gamma_1+\gamma_3}\left(\gamma_2\PP_{13}-\gamma_3\PP_{24}\right)
\end{cases}
\end{equation}
%

%§§§§§§§§§
\paragraph{Mandelstam variables}
The Mandelstam variables themselves are
\begin{equation}\label{eq:Mandelstamvariables}
\begin{split}
s&=\frac{\PP_{12}\BPP_{12}}{\gamma_1\gamma_2}+\frac{\PP_{34}\BPP_{34}}{\gamma_3\gamma_4}\\
t&=\frac{\PP_{14}\BPP_{14}}{\gamma_1\gamma_4}+\frac{\PP_{23}\BPP_{23}}{\gamma_2\gamma_3}\\
u&=\frac{\PP_{13}\BPP_{13}}{\gamma_1\gamma_3}+\frac{\PP_{24}\BPP_{24}}{\gamma_2\gamma_4}
\end{split}
\end{equation}
It can be checked that $s+t+u=0$ by momentum conservation.

%-------------------------------------------------------------------------
\subsection{Quartic recombination formula}\label{subsec:Quarticrecombination}
For $n=4$, we get 12 different ways, corresponding to the 12 different ways to chose $p$ and $q$ with $p\neq q$ in $\sum_{i=1}^4\langle p\,i\rangle[i\,q]$, to write the sum
\begin{equation}\label{eq:n4noncons}
\begin{split}
\sum_{i=0}^4\frac{p_i\bar p_i}{\gamma_i}&=-\frac{\PP_{12}\BPP_{23}}{\gamma_1\gamma_2\gamma_3}-\frac{\PP_{14}\BPP_{43}}{\gamma_1\gamma_4\gamma_3}=-\frac{\PP_{23}\BPP_{34}}{\gamma_2\gamma_3\gamma_4}-\frac{\PP_{21}\BPP_{14}}{\gamma_2\gamma_1\gamma_4}\\
&=-\frac{\PP_{34}\BPP_{41}}{\gamma_3\gamma_4\gamma_1}-\frac{\PP_{32}\BPP_{21}}{\gamma_3\gamma_2\gamma_1}=-\frac{\PP_{41}\BPP_{12}}{\gamma_4\gamma_1\gamma_2}-\frac{\PP_{43}\BPP_{32}}{\gamma_4\gamma_3\gamma_2}\\
&=-\frac{\PP_{13}\BPP_{32}}{\gamma_1\gamma_3\gamma_2}-\frac{\PP_{14}\BPP_{42}}{\gamma_1\gamma_4\gamma_2}=-\frac{\PP_{24}\BPP_{43}}{\gamma_2\gamma_4\gamma_3}-\frac{\PP_{21}\BPP_{13}}{\gamma_2\gamma_1\gamma_3}\\
&=-\frac{\PP_{31}\BPP_{14}}{\gamma_3\gamma_1\gamma_4}-\frac{\PP_{32}\BPP_{24}}{\gamma_3\gamma_2\gamma_4}=-\frac{\PP_{42}\BPP_{21}}{\gamma_4\gamma_2\gamma_1}-\frac{\PP_{43}\BPP_{31}}{\gamma_4\gamma_3\gamma_1}\\
&=-\frac{\PP_{12}\BPP_{24}}{\gamma_1\gamma_2\gamma_4}-\frac{\PP_{13}\BPP_{34}}{\gamma_1\gamma_3\gamma_4}=-\frac{\PP_{23}\BPP_{31}}{\gamma_2\gamma_3\gamma_1}-\frac{\PP_{24}\BPP_{41}}{\gamma_2\gamma_4\gamma_1}\\
&=-\frac{\PP_{34}\BPP_{42}}{\gamma_3\gamma_4\gamma_2}-\frac{\PP_{31}\BPP_{12}}{\gamma_3\gamma_1\gamma_2}=-\frac{\PP_{41}\BPP_{13}}{\gamma_4\gamma_1\gamma_3}-\frac{\PP_{42}\BPP_{23}}{\gamma_4\gamma_2\gamma_3}
\end{split}
\end{equation}

The idea is now to use this information to reverse engineer the sum over the free hamiltonians into an expansion over an overcomplete basis $\{\PP_{12},\PP_{13},\PP_{14},\PP_{23},\PP_{24},\PP_{34}\}$. For that, make the ansatz
\begin{equation}
\begin{split}
\sum_{r=0}^4\frac{p_r\bar p_r}{\gamma_r}&=-\Big(\sum_{r=0}^4K^{12}_r\frac{\bar{p}_r}{\gamma_r}\PP_{12}+\sum_{r=0}^4K^{34}_r\frac{\bar{p}_r}{\gamma_r}\PP_{34}\\
&+\sum_{r=0}^4K^{14}_r\frac{\bar{p}_r}{\gamma_r}\PP_{14}+\sum_{r=0}^4K^{23}_r\frac{\bar{p}_r}{\gamma_r}\PP_{23}\\
&+\sum_{r=0}^4K^{13}_r\frac{\bar{p}_r}{\gamma_r}\PP_{13}+\sum_{r=0}^4K^{24}_r\frac{\bar{p}_r}{\gamma_r}\PP_{24}\Big)
\end{split}
\end{equation}

In \eqref{eq:n4noncons} we can find four terms multiplying $\PP_{12}$. Taking all four and comparing to the first term in the ansatz, we can identify the recombination coefficients $K^{12}_r$. Doing likewise for the rest of the overcomplete basis vectors, 
we find the following solution.
\begin{table}[h]
\centering
\begin{tabular}{c|cccc}
\toprule
$K^{ij}_r$ & $r=1$ & $r=2$ & $r=3$ & $r=4$\\
\midrule
$12$ & $0$ & $0$ & $\frac{1}{\gamma_1}-\frac{1}{\gamma_2}$ & $\frac{1}{\gamma_1}-\frac{1}{\gamma_2}$\\
$34$ & $\frac{1}{\gamma_3}-\frac{1}{\gamma_4}$ & $\frac{1}{\gamma_3}-\frac{1}{\gamma_4}$ & $0$ & $0$\\
\midrule
$14$ & $0$ & $\frac{1}{\gamma_1}-\frac{1}{\gamma_4}$ & $\frac{1}{\gamma_1}-\frac{1}{\gamma_4}$ & $0$\\
$23$ & $\frac{1}{\gamma_2}-\frac{1}{\gamma_3}$ & $0$ & $0$ & $\frac{1}{\gamma_2}-\frac{1}{\gamma_3}$\\
\midrule
$13$ & $0$ & $\frac{1}{\gamma_1}-\frac{1}{\gamma_3}$ & $0$ & $\frac{1}{\gamma_1}-\frac{1}{\gamma_3}$\\
$24$ & $\frac{1}{\gamma_2}-\frac{1}{\gamma_4}$ & $0$ & $\frac{1}{\gamma_2}-\frac{1}{\gamma_4}$ & $0$\\
\bottomrule
\end{tabular} 
\caption{n=4 recombination coefficients.}
\label{tab:N4Coeffs}
\end{table}

It then turns out that these coefficients solves the recombination problem also for arbitrary $c_r$. Indeed we get
\begin{equation}\label{eq:recombinationFormulaQuartic1}
\begin{split}
\sum_{r=1}^4 c_rp_r&=\frac{1}{4}\Big(\sum_{r=1}^4 c_r\gamma_r\Big)\Big(\sum_{s=1}^4\frac{p_s}{\gamma_s}\Big)\\
&-\frac{1}{4}\Big(\sum_{r=0}^4K^{12}_rc_r\PP_{12}+\sum_{r=0}^4K^{34}_rc_r\PP_{34}\\
&+\sum_{r=0}^4K^{14}_rc_r\PP_{14}+\sum_{r=0}^4K^{23}_rc_r\PP_{23}\\
&+\sum_{r=0}^4K^{13}_rc_r\PP_{13}+\sum_{r=0}^4K^{24}_rc_r\PP_{24}\Big)
\end{split}
\end{equation}
or explicitly
\begin{equation}\label{eq:recombinationFormulaQuartic2}
\begin{split}
\sum_{r=1}^4 c_rp_r&=\frac{1}{4}\Big(\sum_{r=1}^4 c_r\gamma_r\Big)\Big(\sum_{s=1}^4\frac{p_s}{\gamma_s}\Big)\\
-\frac{1}{4}\Big(&(c_3+c_4)\frac{\gamma_1-\gamma_2}{\gamma_1\gamma_2}\PP_{12}+(c_1+c_2)\frac{\gamma_3-\gamma_4}{\gamma_3\gamma_4}\PP_{34}\\
+&(c_2+c_3)\frac{\gamma_1-\gamma_4}{\gamma_1\gamma_4}\PP_{14}+(c_1+c_4)\frac{\gamma_2-\gamma_3}{\gamma_2\gamma_3}\PP_{23}\\
+&(c_2+c_4)\frac{\gamma_1-\gamma_3}{\gamma_1\gamma_3}\PP_{13}+(c_1+c_3)\frac{\gamma_2-\gamma_4}{\gamma_2\gamma_4}\PP_{24}\Big)
\end{split}
\end{equation}
This is the quartic recombination formula written in an overcomplete basis. As such it naturally generalises the cubic formula if that one is also written in an overcomplete basis  (look forward to formula \eqref{eq:recombinationFormulaQubic2}).

Next we turn to representing the formula in an independent basis.

%§§§§§§§§§
\paragraph{s-channel representation}
We can use the equations \eqref{eq:SchannelSolution} to express the sum in a $\{\PP_{12},\PP_{34}\}$ basis
\begin{equation}\label{eq:recombinationFormulaQuartic3Schannel}
\begin{split}
\sum_{r=1}^4 c_rp_r=&\frac{1}{4}\Big(\sum_{r=1}^4 c_r\gamma_r\Big)\Big(\sum_{s=1}^4\frac{p_s}{\gamma_s}\Big)\\
-&\frac{1}{4}\left(S_{12,1}c_1+S_{12,2}c_2+S_{12,3}c_3+S_{12,4}c_4\right)\PP_{12}\\
-&\frac{1}{4}\left(S_{34,1}c_1+S_{34,2}c_2+S_{34,3}c_3+S_{34,4}c_4\right)\PP_{34}
\end{split}
\end{equation}
where the coefficients are rational functions of the $\gamma$'s
\begin{align}
S_{12,1}&=\frac{3\gamma_2+\gamma_1}{\gamma_2(\gamma_1+\gamma_2)} & S_{34,3}&=\frac{3\gamma_4+\gamma_3}{\gamma_4(\gamma_3+\gamma_4)}\\
S_{12,2}&=-\frac{3\gamma_1+\gamma_2}{\gamma_1(\gamma_1+\gamma_2)} & S_{34,4}&=-\frac{3\gamma_3+\gamma_4}{\gamma_3(\gamma_3+\gamma_4)}\\
S_{12,3}&=\frac{\gamma_3(\gamma_1-\gamma_2)}{\gamma_1\gamma_2(\gamma_1+\gamma_2)}& S_{34,1}&=\frac{\gamma_1(\gamma_3-\gamma_4)}{\gamma_3\gamma_4(\gamma_3+\gamma_4)}\\
S_{12,4}&=\frac{\gamma_4(\gamma_1-\gamma_2)}{\gamma_1\gamma_2(\gamma_1+\gamma_2)}& S_{34,2}&=\frac{\gamma_2(\gamma_3-\gamma_4)}{\gamma_3\gamma_4(\gamma_3+\gamma_4)}
\end{align}
The coefficients are listed so that it is easy to see that the formula is symmetric under the interchange of labels $1\leftrightarrow 3$ and $2\leftrightarrow 4$. Similar formulas can be written for $\mathbf{t}$-channel and $\mathbf{u}$-channel variables.

%====================================================
\section{Quintic and higher order}\label{sec:QuinticHigher}
%====================================================
In retrospect we can now rewrite the cubic formula in the same way. The formulas \eqref{eq:DefBlackboardMomentumExplicitN3} and \eqref{eq:recombinationFormulaCubic2ndterm} can be used to write 
\begin{equation}\label{eq:recombinationFormulaQubic2}
\begin{split}
\sum_{r=1}^3 c_rp_r&=\frac{1}{3}\Big(\sum_{r=1}^3 c_r\gamma_r\Big)\Big(\sum_{s=1}^3\frac{p_s}{\gamma_s}\Big)\\
-\frac{1}{3}\Big(&c_1\frac{\gamma_2-\gamma_3}{\gamma_2\gamma_3}\PP_{23}+c_2\frac{\gamma_1-\gamma_3}{\gamma_1\gamma_3}\PP_{13}+c_3\frac{\gamma_1-\gamma_2}{\gamma_1\gamma_2}\PP_{12}\Big)
\end{split}
\end{equation}

The table of coefficients, analogous to Table \ref{tab:N4Coeffs}, become Table \ref{tab:N3Coeffs}.

\begin{table}[h]
\centering
\begin{tabular}{c|ccc}
\toprule
$K^{ij}_r$ & $r=1$ & $r=2$ & $r=3$ \\
\midrule
$23$ & $\frac{1}{\gamma_2}-\frac{1}{\gamma_3}$ & $0$ & $0$ \\
$13$ & $0$ & $\frac{1}{\gamma_1}-\frac{1}{\gamma_3}$ & $0$ \\
$12$ & $0$ & $0$ & $\frac{1}{\gamma_1}-\frac{1}{\gamma_2}$ \\
\bottomrule
\end{tabular} 
\caption{n=3 recombination coefficients.}
\label{tab:N3Coeffs}
\end{table}
The tables for $n=3$ and $n=4$ lends themselves to easy generalisation to arbitrary $n$. For $n=5$ we have the overcomplete basis of $10$ transverse momentum variables (of which only three are independent)
\begin{equation*}
\PP_{12},\PP_{13},\PP_{14},\PP_{15},\PP_{23},\PP_{24},\PP_{25},\PP_{34},\PP_{35},\PP_{45}
\end{equation*}
and likewise for the $\BPP_{ij}$.
We can now guess the quintic (n=5) recombination coefficients
\begin{equation}\label{eq:GeneralCoefficients}
\begin{cases}
K^{ij}_r=\frac{1}{\gamma_i}-\frac{1}{\gamma_j}\quad\text{for}\quad r\neq \{i,j\}\cr
K^{ij}_i=K^{ij}_j=0
\end{cases}
\end{equation}
Indeed, this definition of $K^{ij}_r$ works for vertices of arbitrary order $n$.

%§§§§§§§§§
\paragraph{General recombination formula}
With the definitions \eqref{eq:GeneralCoefficients} of the co\-efficients we can write the general order $n$ recombination formula
\begin{equation}\label{eq:recombinationFormulaArbitrary_N}
\begin{split}
\sum_{r=1}^n c_rp_r&=\frac{1}{n}\Big(\sum_{r=1}^n c_r\gamma_r\Big)\Big(\sum_{s=1}^n\frac{p_s}{\gamma_s}\Big)-\frac{1}{n}\sum_{r=1}^n\sum_{i<j}^nK^{ij}_rc_r\PP_{ij}
\end{split}
\end{equation}

It can be checked by direct computation for any $n$ (in practice for small values). It is easiest to do with region variables implementing momentum conservation. A general proof can presumably be constructed by induction over $n$.

%====================================================
\section*{Conclusion}\label{sec:Conclusion}
%====================================================
It is indeed nice to see how naturally the cubic recombination formula generalises to quartic and higher orders.

\pagebreak

\end{document}